\documentclass[twocolumn]{aastex61} 
\usepackage{graphicx}
\usepackage{amsmath}
\usepackage{amssymb,latexsym,amsopn}
\usepackage{color}

\usepackage{ulem}

\shorttitle{FRB dispersions and the circum- and inter-galactic medium}
\shortauthors{Ravi}

\begin{document}

\title{Measuring the circum- and inter-galactic baryon contents with fast radio bursts}

\correspondingauthor{Vikram Ravi}
\email{vikram@caltech.edu}

\author{Vikram Ravi}
\affil{Cahill Center for Astronomy and Astrophysics, MC 249-17, California Institute of Technology, Pasadena, CA 91125, USA.}

\begin{abstract}

Over 80\% of the cosmic baryon density is likely to be distributed in the diffuse, $\gtrsim10^{4}$\,K circum- and inter-galactic medium (CGM and IGM respectively). We demonstrate that the dispersion measures (DMs) of samples of localized Fast Radio Bursts (FRBs) can be used to measure the distribution of baryons between the CGM and IGM. We propose to separate the CGM and IGM contributions to FRB DMs by including redshift and mass measurements of intervening galaxies. Using simulated samples of FRB sightlines through intervening galaxy halos and an illustrative model for the CGM, and including realistic observational uncertainties, we show that small samples ($O(10^{1})-O(10^{2})$) of localized FRBs are sensitive to the presence of CGM gas. The fractions of baryons in the CGM and IGM can be accurately estimated with 100 FRBs at $z<1$, and the characteristic radial density profiles of CGM halos may also be possible to constrain. The required samples of localized FRBs are expected to be assembled in the coming few years by instruments such as the Australian Square Kilometre Array Pathfinder, the Very Large Array, and the Deep Synoptic Array. 

\end{abstract}


\section{Introduction} \label{sec:intro}

Up to $10\%$ of the cosmic baryon fraction, $\Omega_{b}$, is to be found in stars and the interstellar medium \citep{fp04}. The remainder is distributed between the circum-galactic medium \citep[CGM;][]{tpw17}, and the filaments of the intergalactic medium \citep[IGM;][]{m16}. However, in the redshift $z\sim0$ Universe, the bulk of the CGM ($\gtrsim99\%$ by mass) and the IGM ($\gtrsim90\%$) is at temperatures $>10^{4}$\,K and therefore largely ionized, making it difficult to observe. Quasar absorption line studies in the rest-frame ultraviolet, probing HI and ionized-metal transitions corresponding to collisional- and photo-ionization characteristic temperatures up to $10^{6}$\,K, paint a picture of largely mixed, multiphase, kinematically complex CGM/IGM gas. The hottest ($>10^{6}$\,K) gas has been detected as extended X-ray thermal halos around nearby galaxies \citep[e.g.,][]{abd13}, and through the thermal Sunyaev-Z'eldovich effect in IGM filaments \citep{dch+17,thm+17}. However, these studies rely on careful modeling of the density, temperature and chemical profiles of the gas to derive total gas contents, making the overall fractions of $\Omega_{b}$ in the IGM ($f_{\rm IGM}$) and CGM ($\sim1-f_{\rm IGM}$) highly uncertain \citep[$0.5\lesssim f_{\rm IGM}\lesssim0.9$;][]{ssd12,wpt+14}. The dependence of CGM mass on halo mass is also poorly constrained by observations, but may form a crucial discriminant between models for thermal and kinetic feedback in galaxies \citep{srk+17,fqm+17,wds+17}. Further, the characteristic radial density profile of the CGM is highly uncertain, although there are unsurprising indications that it is flatter in shape than the isothermal-sphere case \citep{acb16,pww+17}.

Understanding the distribution of baryons within and between the CGM and IGM is a fundamental astronomical problem, with critical implications for the growth mechanisms of galaxies from extragalactic gas. Here we consider whether detailed observations of Fast Radio Burst (FRB) sightlines can be used to measure (\textit{a}) the CGM/IGM baryon fractions (parameterized by $f_{\rm IGM}$), and (\textit{b}) the radial density profiles of the CGM ($\rho_{\rm CGM}(r)$, where $r$ is the galactocentric radius). FRBs are extragalactic GHz-frequency events of $\mu$s-ms durations, characterized by delays due to dispersion in intervening free-electron columns well in excess of Galactic expectations for their sightlines. FRBs are found at extragalactic dispersion measures (DMs ${\rm DM}_{E}$) of between $150-2600$\,pc\,cm$^{-3}$. If these DMs are modeled as primarily arising in the IGM, FRB redshifts of between $0.18-2$ are suggested \citep{dgb+15,sd18}. However, non-negligible contributions to ${\rm DM}_{E}$ from FRB host galaxies are also possible, and indeed favored by some FRB models \citep[e.g.,][]{kon15,csp06,yll+17,wmb18}. In the case of the repeating FRB\,121102 (${\rm DM}_{E}\approx340$\,pc\,cm$^{-3}$) localized to a star-forming region of a $z=0.193$ dwarf galaxy \citep{clw+17,tbc+17}, the host DM contribution was limited to be $\lesssim250$\,pc\,cm$^{-3}$, assuming that the associated H$\alpha$-emitting nebula traced the entire host DM \citep{kms+17}. Other FRBs are unlikely to originate in magneto-ionic environments as extreme as that of FRB\,121102 \citep{bta+17,msh+18}. For example, the sparsely populated localization region of the ultra-bright FRB\,150807 (${\rm DM}_{E}\approx200$\,pc\,cm$^{-3}$) suggested a distance in excess of 500\,Mpc, and its low Faraday rotation measure in comparison with its scattering properties suggested a host interstellar medium unlike even that of the Milky Way \citep{rsb+16}. 

We focus on the prospects for FRBs that are localized to individual host galaxies, such that host- and intervening-galaxy redshift measurements are possible. Samples of a few hundred FRBs localized with sufficient \citep[few-arcsecond;][]{eb17} accuracy upon the first instances of their detection are expected in the coming few years from the Australian Square Kilometre Array Pathfinder \citep[ASKAP;][]{bsm+17}, the {\em realfast} system at the Jansky Very Large Array \citep[VLA;][]{lbb+18}, and the Deep Synoptic Array (DSA; Ravi et al., in prep.). Although these surveys are well motivated by the problem of FRB progenitors, we argue that they may further result in impactful insights into the CGM/IGM. Our work builds on previous studies of similar intent by \citet{m14}, \citet{dz14}, and \citet{zok+14}. However, our approach is distinct from these works in that we consider what may be achieved with redshift measurements of FRB host galaxies together with redshift and mass estimates for a sample of intervening galaxies. Our simulations of samples of FRBs and intervening galaxy halos are described in \S\ref{sec:model}. The aim of the simulations is to ascertain whether a sample of $N_{\rm FRB}$ FRBs, each with host- and intervening-galaxy  measurements, can be used to estimate $f_{\rm IGM}$ and $\rho_{\rm CGM}(r)$. We hypothesize that this can be done by comparing measurements of the summed CGM and IGM components of FRB DMs, ${\rm DM}_{\rm EG}$, with predictions given the redshifts and masses of intervening galaxy halos and the FRB redshifts. We demonstrate the potential of this technique with simulations of observed FRB samples in \S\ref{sec:res}, and summarize and discuss our results in \S\ref{sec:conc}. We adopt the latest Planck cosmological parameters \citep{pc16}, with $H_{0}=67.7$\,km\,s$^{-1}$\,Mpc$^{-1}$, $\Omega_{b}=0.0486$, $\Omega_{M}=0.3089$, $\Omega_{\Lambda}=0.6911$, and $\sigma_{8}=0.8159$.

\section{The simulation} \label{sec:model}

We begin by assembling a sample of FRB redshifts. The lack of FRB redshift measurements besides FRB\,121102, combined with the uncertain relation between DM and redshift and unknown characteristic host-DM contributions, means that it is difficult to motivate a specific FRB redshift distribution. For our simulations, we therefore adopt an undemanding model for cosmological FRBs wherein (\textit{a}) the FRB volumetric rate traces the star-formation rate, motivated by the bevy of progenitor models that favor young compact objects; (\textit{b}) the FRB luminosity function has the form $N(>L)\propto L^{-0.7}$, motivated by observations of FRB\,121102 \citep{lab+17}; and (\textit{c}) the intrinsic FRB radiation spectrum is flat \citep{gsp+18}. Then, the differential FRB rate above some detection threshold, $\frac{dR}{dz}$, is given by
\begin{equation}
    \frac{dR}{dz} \propto \frac{4\pi d^{2}V_{c}}{d\Omega dz}\rho_{*}(z)D_{L}^{-1.4}(z),
    \label{eqn:1}
\end{equation}
where $\frac{d^{2}V_{c}}{d\Omega dz}$ is the standard differential comoving volume element, $D_{L}(z)$ is the luminosity distance, and we adopt the fit to the cosmic star-formation rate density, $\rho_{*}(z)$, of \citet{hb06} (their Table~1, final column).  

\begin{figure}[h]
\centering
\includegraphics[scale=0.44,angle=-90]{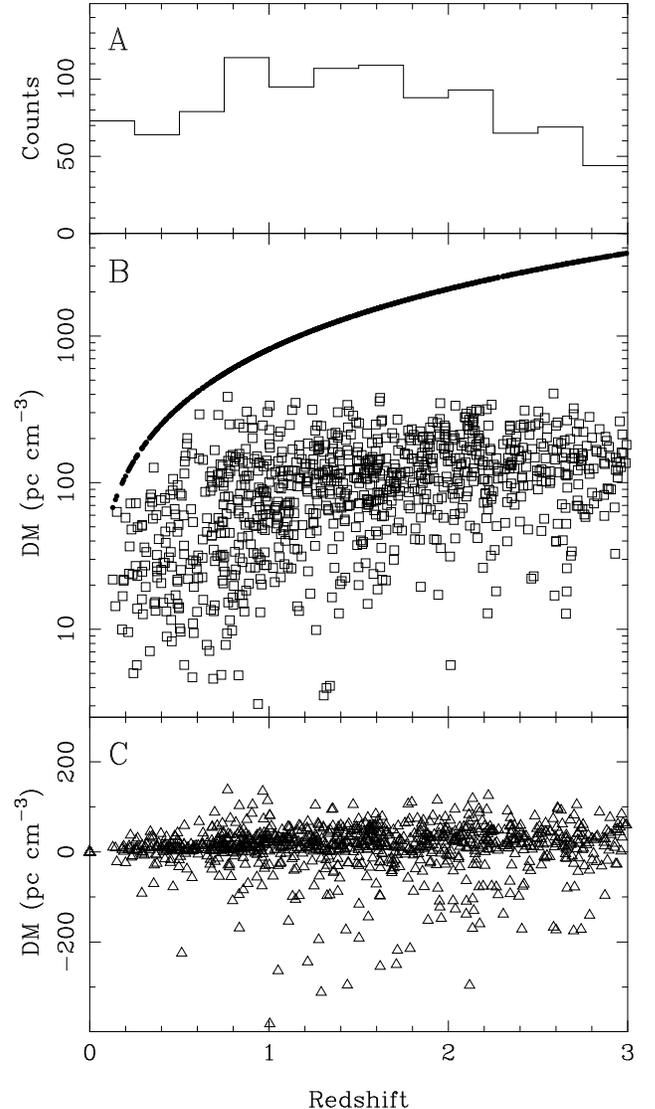}
\caption{{\em Panel A:} Histogram of 1000 simulated FRB redshifts, randomly drawn according to Equation~\ref{eqn:1}. {\em Panel B:} Contributions to FRB DMs from the IGM (solid points; no scatter included) and CGM (open squares) for each simulated FRB sightline. We adopted the Model~A (constant-density) CGM radial density profile, and assumed $f_{\rm IGM}=0.5$. Note that the results for other values of $f_{\rm IGM}$ can be derived through a straightforward linear scaling by $f_{\rm IGM}/0.5$. {\em Panel C:} Difference between the DM-contributions from the CGM for Model A and B (isothermal-sphere) CGM density profiles. Model~A profiles result in typically larger CGM DM contributions.}
\label{fig:1}
\end{figure}

Next, for an FRB at a redshift $z_{\rm FRB}$, we specify the distribution of intervening galaxy halos in their mass, $M_{h}$, and $z$. We approximate the extent of each halo by its (approximate) virial radius, $r_{\rm 200}$ \citep{cl96}, and adopt an NFW \citep{nfw96} halo density profile and \citet{dsk+08} concentration parameters to calculate $r_{\rm 200}(M_{h},z)$. The halo mass function, $dn(M_{h},z)/dM_{h}$, is in turn specified according to \citet{smt01}, as implemented by \citet{mpr13}. Then, the differential number of halo intercepts is given by \citep[e.g.,][Equation~31]{h99}
\begin{equation}
    \frac{d^{2}N}{dM_{h}dz} = 4\pi r_{200}^{2}(M_{h},z)\frac{dn(M_{h},z)}{dM_{h}}\frac{c(1+z)^{2}}{H(z)},
    \label{eqn:2}
\end{equation}
where $c$ is the vacuum speed of light, and $H(z)$ is the Hubble parameter. For each FRB redshift, we use this distribution function to draw a sample of intervening halos. Throughout this work, we only consider halo masses in the range $10^{11}M_{\odot}$ to $10^{15}M_{\odot}$. The CGM contents of lower-mass halos, corresponding to stellar masses $\lesssim10^{9}M_{\odot}$ \citep{bcw10}, are unlikely to be maintained in thermal equilibrium by virial shocks, and are therefore strongly influenced by galactic feedback mechanisms \citep{fqm+17}. Many simulations \citep[e.g.,][]{sfb+15,wds+17} find that feedback in low-mass halos results in smaller CGM mass fractions than in higher-mass halos \citep[although see][]{srk+17}, in tentative agreement with observations \citep{tpw17}. Further, the baryonic components of halos with masses $<10^{10}M_{\odot}$ are below the IGM Jeans mass, and are unlikely to have collapsed. These arguments, together with the statistics of halo intercepts specified by Equation~\ref{eqn:2} suggests that the contributions of the CGM in $M_{h}<10^{11}M_{\odot}$ halos to FRB DMs may not be significant. 

We consider two illustrative boundary-case models for $\rho_{\rm CGM}(r)$ to calculate the DM contributions from each halo. In Model~A, we assume a constant-density CGM at $r<r_{\rm 200}$. In Model~B, we assume an isothermal sphere truncated at $r_{\rm 200}$, such that $\rho_{\rm CGM}(r)\propto r^{-2}$.  These cases bound what is measured \citep{acb16,pww+17}, and results from simulations \citep{fqm+17} that suggest $\rho_{\rm CGM}(r)\propto r^{-1.5}$. At each redshift, we normalize the radial density profiles by requiring that the total CGM mass of each halo be given by $(1-f_{\rm IGM})M_{h}\Omega_{b}/\Omega_{M>11}$, where $\Omega_{M>11}$ is the fraction of the critical density in $M_{h}>10^{11}M_{\odot}$ halos. We neglect the fraction of $\Omega_{b}$ in stars and the interstellar medium ; this is justified because our analysis is agnostic to the actual fractions of $\Omega_{b}$ in the CGM and IGM. The electron densities are calculated following \citet{sd18}, with the assumption of no significant difference in the ionization fractions of the CGM and IGM. The impact parameters of the FRB sightlines with respect to the intervening halos are drawn from a distribution proportional to $r^{-1}$, and we correct the DM contributions of each halo by a factor $(1+z)^{-1}$ to account for the redshifting of FRB emission. Finally, to specify the IGM contributions to ${\rm DM}_{\rm EG}$, we adopt the formalism of \citet{sd18} for the DM contribution from a constant-density IGM (their equations 4 and 5, but corrected to include a factor of $(1+z)^{-1}$ in the integrand of equation 5). \citet{sd18} suggest an intrinsic scatter of $\sigma_{\rm IGM}\approx10$\,pc\,cm$^{-3}$, accounting for cosmic-web voids and filaments.

In Fig.~\ref{fig:1}, we show simulations of the CGM and IGM contributions to ${\rm DM}_{\rm EG}$ for 1000 FRB sightlines at various redshifts in the range $z=0-3$. The redshift distribution of the simulated FRBs, specified by Equation~\ref{eqn:1}, is shown in Panel~A. Our assumption of $f_{\rm IGM}=0.5$ in the figure implies typical CGM DM contributions of a few hundred pc\,cm$^{-3}$ for FRBs beyond $z=1$ (Panel~B), while the fractional contribution of the CGM to ${\rm DM}_{\rm EG}$ is largest for lower redshifts. For the purposes of illustration, no scatter has been included in the IGM DMs in Panel~B. Panel~C indicates that Model~A CGM radial density profiles (constant-density) result in typically larger CGM DMs than Model~B profiles (isothermal sphere), which is expected given the greater mass concentration in Model~B. 

\section{Results} \label{sec:res}

We use the simulations described above to ascertain whether $f_{\rm IGM}$ and $\rho_{\rm CGM}(r)$ can be estimated using samples of localized FRBs. The method we propose is to compare measurements of the summed CGM$+$IGM FRB DMs, $\widehat{{\rm DM}}_{\rm EG}$, with predictions for ${\rm DM}_{\rm EG}$. The predictions, which are based on redshift and mass measurements of intervening galaxies, and measurements of FRB redshifts, depend on an assumed $f_{\rm IGM}$ to partition free electrons between the CGM and IGM, and on an assumed $\rho_{\rm CGM}(r)$ to calculate the DM contributions from each intervening galaxy halo. Thus, measurements of $\widehat{{\rm DM}}_{\rm EG}$ will only be consistent with predictions of ${\rm DM}_{\rm EG}$ for a unique combination of $f_{\rm IGM}$ and $\rho_{\rm CGM}(r)$. 

We first consider how samples of FRBs with identified intervening galaxies are assembled in practice. Potential sources of error in both the estimates of and predictions for the CGM/IGM DMs are assessed. We then demonstrate the effects of these errors on estimates of $f_{\rm IGM}$ and $\rho_{\rm CGM}(r)$ using realistic samples of localized FRBs. 

\subsection{Observational considerations} \label{sec:obs}

Constructing the estimate $\widehat{{\rm DM}}_{\rm EG}$ for an observed FRB relies on subtracting all other contributions from the measured DM. Each subtraction has a corresponding uncertainty. First, FRB DM contributions from the Milky Way disk are traditionally estimated by integrating the NE2001 model for the warm ionized medium density structure \citep{cl02} to its outer edge, resulting in values of $\sim30/\sin|b|$\,pc\,cm$^{-3}$ at high Galactic latitudes $b$. Negligible uncertainty is expected in these estimates for $|b|\gtrsim20$\,deg \citep{gmc+08,dgb+15}. Next, the Milky Way hot halo (i.e., its CGM) is expected to produce $\sim40$\,pc\,cm$^{-3}$ of DM for every FRB, with an uncertainty of $\sigma_{\rm MW}\approx15$\,pc\,cm$^{-3}$ \citep{dgb+15}. DM contributions from FRB host galaxies are highly uncertain, and dependent on specific progenitor models \citep{xh15,yll+17,wmb18}. Here we assume that as larger  samples of localized, thoroughly characterized FRBs are constructed, further insight into FRB progenitors will be gleaned from, for example, their host galaxies, positions with respect to their hosts, characteristic luminosities, spectra, polarizations, scattering and Faraday-rotation properties, repeatability, and potential multiwavelength counterparts. Further, given that exceedingly large host DM contributions are likely excluded in some known cases \citep{rsb+16}, we assume that host-galaxy DMs can be subtracted with a conservative uncertainty of $\sigma_{\rm host}\approx50$\,pc\,cm$^{-3}$ \citep[e.g., Fig.~3 of][]{wmb18}. Adding $\sigma_{\rm MW}$, $\sigma_{\rm host}$ and $\sigma_{\rm IGM}\approx10$\,pc\,cm$^{-3}$ (see above) in quadrature results in an uncertainty of $\sigma_{\rm EG}\approx53$\,pc\,cm$^{-3}$.

\begin{figure}[h]
\centering
\includegraphics[scale=0.64,angle=-90]{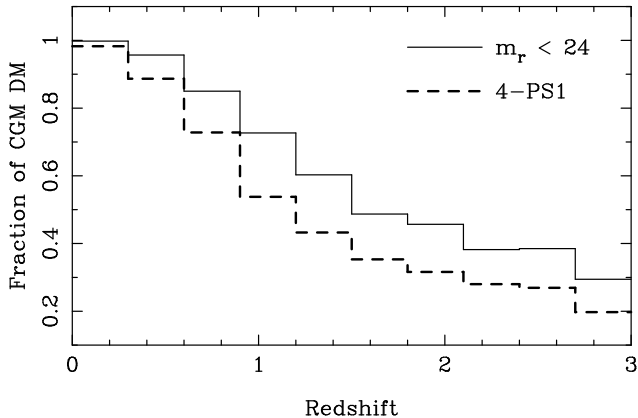}
\caption{This figure demonstrates the impacts of observational limitations in identifying intervening galaxies on estimates of CGM DM contributions to ${\rm DM}_{\rm EG}$. The curves show the mean fractions of FRB CGM DMs contributed by observed intervening galaxies in ten bins of redshift. Two observational schemes are considered: the case where galaxies are identified with apparent $r$-band magnitudes $m_{r}<24$ (solid curve), and the case where detections in four of the five Pan-STARRS1 $3\pi$-survey stack filters are required \citep{cmm+16} (dashed curve). Although Model~A (constant-density) CGM density profiles were assumed, adopting Model~B profiles does not significantly alter the results.}
\label{fig:2}
\end{figure}

Predicting ${\rm DM}_{\rm EG}$ for an FRB sightline relies on observationally identifying intervening galaxy halos, and measuring their redshifts and masses. We consider a scheme whereby candidate intervening galaxies are identified through optical/IR imaging, perhaps including color information to estimate photometric redshifts, and spectra are obtained using multi-object spectrographs to confirm redshifts. Intervening galaxies widely separated from FRB sightlines are unlikely: for example, the sample of 1000 FRB sightlines presented in Fig.~\ref{fig:1} contains only seven intervening galaxies with projected offsets $>10\arcmin$, with a maximum offset of 16.4\arcmin, out of 4161 intervening galaxies. To assess the completeness of imaging observations of specific depths to intervening galaxies, we assign optical/IR spectral energy distributions (SEDs) to simulated intervening dark-matter halos using the publicly available output catalogs recent semi-analytic galaxy formation model \citep{hwt+15}. We obtained rest-frame dust-corrected SEDs between the GALEX-FUV and K bands for halos in the mass range $10^{11}-10^{15}M_{\odot}$ for each redshift snapshot, and binned them in 0.04\,dex $M_{h}$-bins. For each simulated intervening halo, we then randomly drew an SED from the nearest mass- and redshift-bin, $K$-correcting the observed SED and accounting for the halo luminosity distance. We consider two means of selecting candidate intervening systems for spectroscopic follow-up: detection in four of the five filters of the Pan-STARRS1 $3\pi$ survey stack \citep[PS1;][]{cmm+16}, and detection in an $r$-band image with $m_{r}<24$ (AB). 

In Fig.~\ref{fig:2}, we show the impact of these observational selections on the typical completeness of intervening galaxy samples for FRBs at various redshifts. We present the mean fractions of the total CGM DMs for FRBs in ten redshift bins in the range $z=0-3$ recovered by the two observational selections. For example, for a $z<1$ FRB, $>50\%$ of the CGM DM is expected on average to be contributed by galaxies detected in four PS1 filters. For a $z<0.5$ FRB, $r$-band imaging observations with a limiting magnitude of $m_{r}=24$ will detect galaxies contributing on average $>90\%$ of the CGM DM. 

In estimating dark-matter halo masses, the tight relation between halo and stellar masses \citep[intrinsic scatter $\approx0.16$\,dex;][]{bcw10} implies that stellar-mass estimation errors can predominantly contribute to halo mass errors. Based on the assessment of the stellar-mass estimation error budget by \citet{mdf+15}, we consider stellar-mass estimation errors of 0.25\,dex. In particular, we assume that deep follow-up imaging renders photometric errors negligible, and that the existence of spectroscopic data on each galaxy enables the accurate modeling of nebular emission lines. Thus, our total scatter in halo-mass estimates is 0.3\,dex.

\subsection{Estimating $f_{\rm IGM}$ and $\rho_{\rm CGM}(r)$}

\begin{figure*}[!ht]
\centering
\includegraphics[scale=0.4]{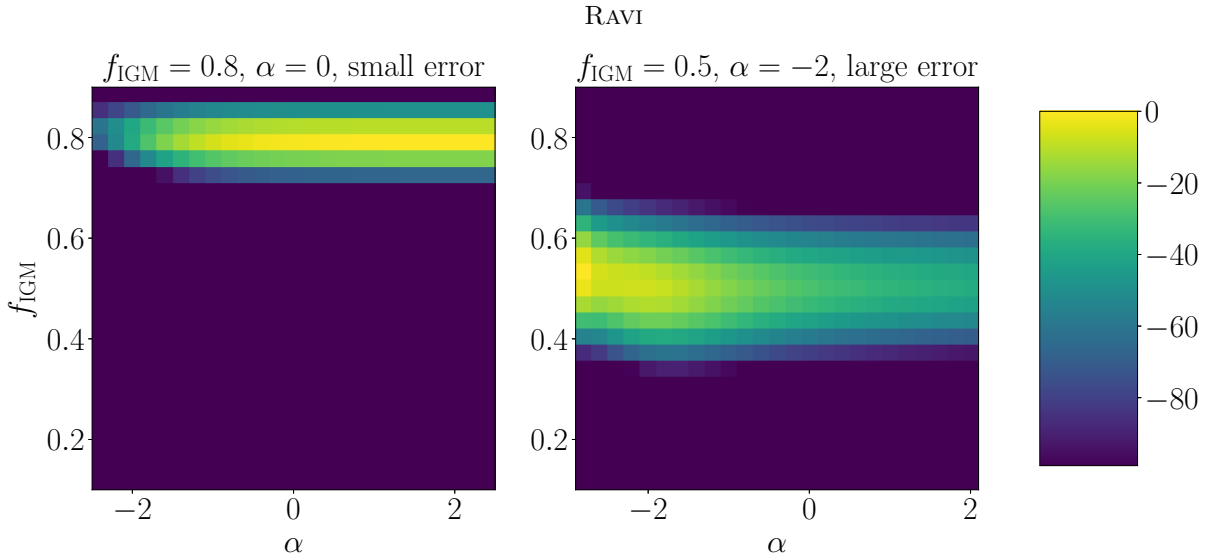}
\caption{Each panel depicts the relative (natural) log-likelihood in the $f_{\rm IGM}-\alpha$ plane for individual simulated samples of 100 FRBs. \textit{Left: } Case~1 sample (see text for details). \textit{Right:} Case~2 sample. The true values of $f_{\rm IGM}$ and $\alpha$ are indicated above each panel.}
\label{fig:3}
\end{figure*}

We consider a maximum-likelihood estimate of $f_{\rm IGM}$ and $\rho_{\rm CGM}(r)$ given a sample of FRB sightlines, where each FRB $i$ is accompanied by measurements of $\widehat{{\rm DM}}^{i}_{\rm EG}$, and predictions of ${\rm DM}^{i}_{\rm EG}(f_{\rm IGM},\alpha)$. For the purposes of estimation, we adopt the parameterization $\rho_{\rm CGM}(r)\propto r^{\alpha}$. Assuming normally distributed errors with variance $\sigma_{\rm DM}^{2}$, the likelihood function is specified by 
\begin{equation}
    \mathcal{L}(f_{\rm IGM},\alpha) \propto \prod_{i}\exp[-(\widehat{{\rm DM}}^{i}_{\rm EG}-{\rm DM}^{i}_{\rm EG})^2/(2\sigma_{\rm DM}^{2})].
\end{equation}
We demonstrate the estimation of $f_{\rm IGM}$ and $\alpha$ by numerically evaluating this likelihood function in two cases:
\begin{description}
    \item[Case~1] True values of $f_{\rm IGM}=0.8$ and $\alpha=0$ (Model~A CGM density profiles). Intervening galaxies are first identified in an $r$-band image with a limiting magnitude of $m_{r}=24$. 
    \item[Case~2] True values of $f_{\rm IGM}=0.5$ and $\alpha=-2$ (Model~B CGM density profiles). Intervening galaxies are first identified using detections in four PS1 filters. We also assume that a further 0.3\,dex of uncertainty is combined with the scatter in the CGM density predictions for each intervening galaxy due to potential un-modeled galaxy-to-galaxy variations in CGM mass. 
\end{description}
We evaluate the likelihood for various trial pairs of $f_{\rm IGM}$ and $\alpha$ in each case using simulated FRB samples. In each sample, ``measurements'' $\widehat{{\rm DM}}^{i}_{\rm EG}$ are generated by first calculating the true values of the CGM/IGM DMs for each sightline for the assumed $f_{\rm IGM}$ and $\alpha$, and then adding normally distributed error values with zero mean and standard deviation $\sigma_{\rm EG}=53$\,pc\,cm$^{-3}$. For each pair of trial values of $f_{\rm IGM}$ and $\alpha$, predicted CGM contributions to ${\rm DM}_{\rm EG}$ are generated by drawing a random sample of observed intervening galaxies to calculate the CGM DM. We correct each prediction based on the estimated completeness factor for the FRB redshift (as depicted in Fig.~\ref{fig:2}), and include log-normally distributed errors with standard deviations of 0.3\,dex (Case 1 above) and 0.42\,dex (Case 2). Predicted IGM contributions to ${\rm DM}_{\rm EG}$ are calculated as described in \S\ref{sec:model} with no errors added (these errors are absorbed in the $\widehat{{\rm DM}}^{i}_{\rm EG}$ simulations). We estimate the variances $\sigma_{\rm DM}^{2}$ for a given sample by calculating the variance of $\widehat{{\rm DM}}^{i}_{\rm EG}-{\rm DM}^{i}_{\rm EG}$ values, where the predictions ${\rm DM}^{i}_{\rm EG}$ were made with the true values of $f_{\rm IGM}$ and $\alpha$. In both cases, we limit our simulated samples to $z<1$ to ensure reasonable completeness of the intervening-galaxy observations to the CGM DM contributions; most currently observed FRBs are likely to originate from $z<1$ \citep[e.g.,][]{dgb+15,sd18}. 

\begin{figure}[h]
\centering
\includegraphics[scale=0.64,angle=-90]{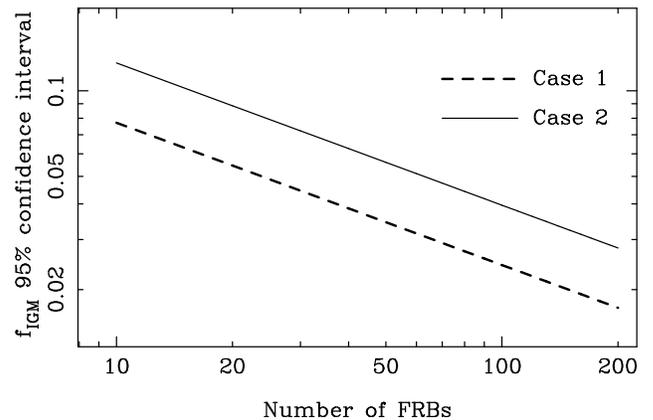}
\caption{The sizes of the 95\% confidence intervals on $f_{\rm IGM}$ in Cases 1 and 2 (see text for details) for FRB samples of different sizes.}
\label{fig:4}
\end{figure}

In Fig.~\ref{fig:3}, we show the relative log-likelihoods in the $f_{\rm IGM}-\alpha$ plane for realizations of 100-FRB samples in Cases 1 and 2. High-significance measurements of $f_{\rm IGM}$ are possible in both cases regardless of the value of $\alpha$, whereas only a weak constraint, equivalent to a lower limit, on $\alpha$ is possible in Case~1. Further simulations that we conducted showed that no useful constraints are possible on $\alpha$ in Case~2. The difference in the $\alpha$-constraints between the cases is due to a combination of the increased uncertainty and lower value of $\alpha$ in Case~2. The greater sensitivity of the technique to $f_{\rm IGM}$ as compared to $\alpha$ is because the variation in the predicted CGM DM with halo impact parameter for different values of $\alpha$ is weaker than the variation in ${\rm DM}_{\rm EG}$ with $f_{\rm IGM}$, within the allowed ranges.

To better quantify the utility of localized FRB samples of different sizes, we refer the reader to Fig.~\ref{fig:4}. Here we plot the sizes of the 95\% confidence intervals on $f_{\rm IGM}$ in Cases 1 and 2 for FRB samples of different sizes, marginalized over $\alpha$. By running simulations in the range $10<N_{\rm FRB}<1000$, we verified that the uncertainty in $f_{\rm IGM}$ is proportional to $N_{\rm FRB}^{-1/2}$. This is not a trivial result, because it depends on whether or not the CGM DM contributions are typically dominated by the largest, rarest intervening halos. This appears not to be the case, as is further indicated by the spread of CGM DM contributions in Fig.~\ref{fig:1}, Panel~B. We find that the uncertainty in $f_{\rm IGM}$ is given by $0.061N_{\rm FRB}^{-1/2}$ in Case~1, and $0.099N_{\rm FRB}^{-1/2}$ in Case~2. These results are highly promising: even in Case~2, a 95\% confidence interval of 0.05 is likely possible with $N_{\rm FRB}=100$.

\section{Summary and discussion} \label{sec:conc}

We present realizations of cosmological FRB sightlines through intervening galaxy halos, with the aim of determining whether samples of localized FRBs are sensitive to the presence of circum-galactic gas. By parameterizing the fractions of $\Omega_{b}$ in the IGM and CGM as $f_{\rm IGM}$ and $1-f_{\rm IGM}$ respectively, and assuming power-law CGM radial density profiles of the form $r^{\alpha}$, we find that  $f_{\rm IGM}$ can be accurately estimated, and weak constraints potentially placed on $\alpha$, with 100 FRBs at $z<1$ (Fig.~\ref{fig:3}). Almost independently of the value of $\alpha$, useful measurements of $f_{\rm IGM}$ can be obtained using samples of $N_{\rm FRB}>10$ localized events (Fig.~\ref{fig:4}). Our work differs from previous studies \citep[e.g.,][]{m14} in that we assume that each FRB is accompanied by a redshift measurement, and that follow-up observations are conducted to measure the redshifts and masses of intervening galaxies (\S\ref{sec:obs}). The initial identification of intervening galaxies in a survey such as the Pan-STARRS $3\pi$ stack is sufficient to recover on average $>50\%$ of our simulated CGM contributions to FRBs at $z<1$ (Fig.~\ref{fig:2}). 

The ASKAP, VLA/{\em realfast}, and DSA surveys are expected to yield a few hundred FRBs localized to individual galaxies in the coming few years, with a significant fraction at $z<1$. However, substantial optical follow-up of each FRB sightline will be required to realize our goal of characterizing the bulk baryon contents of the CGM and IGM. For example, if intervening galaxies were to be selected in deep $r$-band images above a limiting magnitude of $m_{r}=24$, $\sim25$\,arcmin$^{-2}$ galaxies \citep{shy+95} would have to be sifted through in a few$\times$few arcminute region to identify $<10$ intervening galaxies. Initial selections based on photometric redshifts may enable the intervening galaxies to be identified using individual $\sim2$\,hr multi-slit spectroscopic observations with 8-m class telescopes; these will ultimately be available over large areas of the sky from the LSST and DES data sets.  

In practise, analyses such as that we propose may be beset by a selection of systematic uncertainties beyond those included in our simulations. Measurements of the combined CGM and IGM components of FRB DMs ($\widehat{{\rm DM}}_{\rm EG}$) rely on accurate subtraction of other DM components. First, the scatter in host-galaxy DMs may need to be mitigated by the careful selection of FRB samples. For example, it may be necessary to exclude FRBs with similar host environments to the repeating FRB\,121102, for which DMs up to $\sim250$\,pc\,cm$^{-3}$ could be contributed by the host \citep{tbc+17,kms+17}, unless a way to more accurately measure host DMs were found. Even without the $\ll1\arcsec$ localization accuracy required to associate FRB\,121102 with a star-forming region, similar FRBs could be identified by, e.g., the host-galaxy properties or their Faraday-rotation measures.  Second, more scatter than we have assumed may be present in ``IGM'' DMs if, for example, $M_{h}<10^{11}M_{\odot}$ halos retain significant baryon fractions. On the other hand, the statistics of FRB DMs may instead be useful in identifying any unknown sources of DM associated with FRB sightlines, such as dense progenitor environments \citep[e.g.,][]{wmb18}. 

Samples of localized FRBs may provide the best means to determine the distribution of baryons within and between the CGM and IGM. Motivated by the promising results presented here, we will extend this work in a forthcoming paper by analyzing FRB sightlines in cosmological galaxy-formation simulations. Several improvements to our model for the CGM and IGM DMs are desirable, such as: a self-consistent treatment of baryon fractions in stars/dust and multi-temperature gas; the consideration of more sophisticated CGM density structures, extents, and masses that may all vary with galaxy mass and type, and; a robust prescription for baryon density fluctuations outside galaxies, and galaxy clustering. The possibility of FRB observations being affected by and gaining insights into these complexities further motivates the assembly of large samples of localized events.

\acknowledgments

We thank G. Hallinan, P. Hopkins, C. Hummels, and H. Vedantham for useful discussions, and J. Hessels for comments on the manuscript. We made use of the \texttt{astropy} (http://www.astropy.org/), \texttt{hmf} \citep{mpr13}, and \texttt{NFW} \\ (https://github.com/joergdietrich/NFW) Python packages in this work. The Millennium Simulation database used in this paper and the web application providing online access to them were constructed as part of the activities of the German Astrophysical Virtual Observatory.

\end{document}